\documentclass[prl,aps,showpacs,twocolumn,superscriptaddress,amsmath,amssymb]{revtex4}

\usepackage{graphicx,dcolumn,bm}

\begin{document}

\title{Sub-Rayleigh Quantum Imaging Using Single Photon Sources}

\author{C. Thiel}
\affiliation{Institut f\"ur Optik, Information und Photonik, Max-Planck Forschungsgruppe, Universit\"at Erlangen-N\"urnberg, 91058 Erlangen, Germany}
\email{cthiel@optik.uni-erlangen.de}
\homepage{http://www.ioip.mpg.de/jvz/}

\author{T. Bastin}
\affiliation{Institut de Physique Nucl\'eaire, Atomique et de Spectroscopie, Universit\'e de Li\`ege, 4000 Li\`ege, Belgium}

\author{J. von Zanthier}
\affiliation{Institut f\"ur Optik, Information und Photonik, Max-Planck Forschungsgruppe, Universit\"at Erlangen-N\"urnberg, 91058 Erlangen, Germany}

\author{G. S. Agarwal}
\affiliation{Department of Physics, Oklahoma State University, Stillwater, OK 74078-3072, USA}

\date{\today}

\begin{abstract}

We propose a technique capable of imaging a distinct physical object with sub-Rayleigh resolution in an ordinary far-field imaging setup using single-photon sources and linear optical tools only. We exemplify our method for the case of a rectangular aperture and two or four single-photon emitters obtaining a resolution enhanced by a factor of two or four, respectively.

\end{abstract}

\pacs{42.50.St, 42.50.Dv, 03.67-a}

\maketitle

In modern quantum optics, there is a great variety of proposals trying to improve different aspects of the image formation process, commonly summarized by the field of {\em quantum imaging}. Today, this fast growing field ranges from early Ghost imaging~\cite{Pittman:1995:a}, sub-wavelength phase measurements~\cite{Walther:2004:a,Mitchell:2004:a} to quantum lithography~\cite{Boto:2000:a,D'Angelo:2001:a,Hemmer:2006:a}, quantum microscopy~\cite{Teich:1997:a,Muthukrishnan:2004:a,Agarwal:2004:a,Thiel:2007:a} and many more (see e.g.~\cite{Shih:2007:a}). Though all proposals commonly aim to overcome the classical boundaries of image formation, only few improve the spatial resolution itself, i.e.~the ability to image a physical object while overcoming the Rayleigh~\cite{Rayleigh:1879:a} or Abbe limit~\cite{Abbe:1873:a} of classical optics.

So far, in quantum imaging sub-classical resolution has been achieved by using sources of entangled photons~\cite{D'Angelo:2001:a,Muthukrishnan:2004:a}, but it was also shown recently that initially uncorrelated light can be used for that purpose~\cite{Giovannetti:2008:a,Thiel:2007:a}. All those methods exploit second (or $N$th) order correlations between two (or $N)$ photons, i.e.~quantum interferences between two- (or $N\mbox{-})$ photon amplitudes, to surmount the classical boundaries. Hereby, it is yet a challenge to implement sub-Rayleigh quantum imaging using linear optical tools only.

In this letter, we propose a method of imaging a physical object, e.g.~an aperture, beyond the classical Rayleigh resolution using linear optics. Our scheme involves $N$ uncorrelated single photon emitters serving as a non-classical light source and $N$ detectors performing correlation measurements. By placing the $N$ detectors at different positions in the Fourier plane of the object we avoid the use of multiphoton absorption techniques. Moreover, using a lens in the Fourier plane our setup is also capable to reproduce the object in the image plane of the lens. We exemplify our method for the case of two single photon emitters. By exploiting two-photon interferences we show that this scheme allows to image the object with sub-Rayleigh resolution, i.e., with a resolution enhanced by a factor of two with respect to the classical case. In the same way, sub-Rayleigh resolution enhanced by a factor of four is obtained for $N=4$ emitters and using $N=4$ detectors. By extending this scheme, we show that the same results are also obtained for different objects, e.g., in case of a grating with $N$ slits.

Only recently, first experiments were able to observe higher order interferences of photons emitted by single trapped atoms~\cite{Beugnon:2006:a,Maunz:2007:a,Dubin:2007:a,Dubin:2007:b}. These observations stand in a long line of experiments using single photon sources to investigate interference phenomena of single- and multi-photon amplitudes. After the early demonstration of first-oder interferences of light scattered by two ions~\cite{Eichmann:1993:a}, two-photon interferences have been observed and quantified by now in several systems~\cite{Santori:2002:a,Hettich:2002:a,Legero:2004:a,Beugnon:2006:a,Maunz:2007:a,Dubin:2007:a,Dubin:2007:b}. Generally, quantum interferences may appear if a measurement does not allow to reveal information about the particular path a quantum state has propagated (so called {\em Welcher Weg} information). For example, if we detect $N$ ($N>1$) photons by $N$ different detectors placed in the far-field region of a source consisting of $N$ single photon emitters, there are generally $N!$ possibilities how the $N$ photons may propagate from the $N$ atoms to the $N$ detectors~\cite{Thiel:2007:a}. Hereby, the $N!$ quantum paths may differ from each other by an optical phase, leading to destructive or constructive interferences between the $N$-photon amplitudes. This can be fruitfully exploited, e.g., to obtain information about the spatial distribution of the source even if of dimensions smaller than the optical wavelength $\lambda$~\cite{Agarwal:2004:a,Thiel:2007:a}. In the following, we will extend this concept by introducing a distinct physical object between source and detectors which we aim to image with sub-Rayleigh resolution in an ordinary far-field imaging setup.

\begin{figure}[t!]
\centering
\includegraphics[width=0.45\textwidth, bb=65 475 780 810, clip=true]{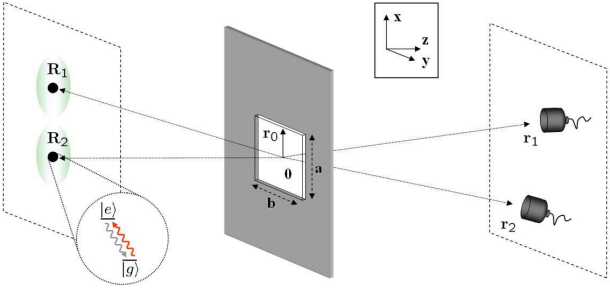}
\caption{\label{fig1} (color online) Setup used for quantum imaging: in a single measurement cycle both emitters at ${\bf R}_1$ and ${\bf R}_2$ each emit a single photon. Between the source and the detection plane, the field is diffracted by an object, e.g.~a rectangular aperture of height $a$ and width $b$. The two detectors are placed in the far-field region of the aperture at positions ${\bf r}_1$ and ${\bf r}_2$.}
\end{figure}

The configuration used in our imaging scheme is shown in Fig.~\ref{fig1}. It consists of two identical two-level atoms at ${\bf R}_1$ and ${\bf R}_2$ initially excited by a single laser $\pi$ pulse (where we denote the initially excited state by $|e_1,e_2\rangle$). After the spontaneous decay of both atoms, the two photons may be recorded by two detectors placed at ${\bf r}_1$ and ${\bf r}_2$ in the far-field region of the source. We next place a physical object between the atoms and the detection plane; in a first step we consider a rectangular aperture with opening height $a$ and width $b$. In a single successful measurement cycle, the two photons emitted by the two atoms pass both by the aperture (distinct to~\cite{Pittman:1995:a}) and are registered by the two detectors. For the sake of simplicity the two atoms and the two detectors are located in the $x$-$y$-plane, parallel to the plane of the aperture, where in particular $R_{1_z}=R_{2_z}:=R_z$ and $r_{1_z}=r_{2_z}:=r_z$. Moreover, we consider coincident detection only~\cite{note1}. Correlating the two detector events we measure the second order correlation function
\begin{eqnarray}\label{G2}
\lefteqn{G^{(2)}({\bf r}_1,{\bf r}_2)=}&\nonumber\\
&\langle e_1,e_2| E^{(-)}({\bf r}_1)E^{(-)}({\bf r}_2)E^{(+)}({\bf r}_2)E^{(+)}({\bf r}_1)|e_1,e_2\rangle,
\end{eqnarray}
which, in the present setup, can be understood as the (unnormalized) joint probability to find one photon at ${\bf r}_1$ and another one at ${\bf r}_2$. Hereby, $E^{(+)}({\bf r}_i)=[E^{(-)}({\bf r}_i)]^\dagger$ denotes the positive frequency part of the electric field amplitude at point ${\bf r}_i$.

From classical diffraction optics (see e.g.~\cite{Born:1999:a}) we know that the electric field $E({\bf r})$ is diffracted at an aperture ${\cal A}$ placed between the source and the detection plane. The disturbed field $U({\bf r})$ can be calculated employing {\em Fresnel-Kirchhoff} diffraction theory. Using standard {\em Fresnel} approximations between source, aperture and detection plane we obtain for the diffracted field at ${\bf r}_i$ the following expression (see~\cite{Born:1999:a})
\begin{eqnarray}\label{Ui}
\hspace{-0.5cm}U({\bf r}_i,{\bf R}_j)=&-&\frac{iA}{\lambda}\;\frac{e^{ik|{R_z-r_{0_z}|}}e^{ik|r_{0_z}-r_z|}}{R_z\,r_z}\\
&\cdot&\int\!\!\!\!\int\limits_{\cal A}e^{i\frac{k}{2}\frac{|{\boldsymbol\rho}_j-{\boldsymbol\rho}_0|^2}{R_z}}\,e^{i\frac{k}{2}\frac{|{\boldsymbol\rho}_0-{\boldsymbol\rho}_i|^2}{r_z}}\;dS({\boldsymbol\rho}_0),\nonumber
\end{eqnarray}
where $A$ denotes the initial amplitude of the electric field, $k=\frac{2\pi}{\lambda}$, ${\bf r}_0$ is a vector in the plane of the aperture (see Fig.~\ref{fig1}) and ${\bf \rho}_\pi$ is a vector consisting of the x- and y-components of ${\bf R}_j$, ${\bf r}_0$ and ${\bf r}_i$, respectively (with $\pi=j,0,i$; $j,i=1,2$). As one can see from Eq.~(\ref{Ui}), the problem is separated into the propagation from the source to the aperture (i.e.~from ${\bf R}_j$ to ${\bf r}_0$) and further from the aperture to the detection plane (i.e.~from ${\bf r}_0$ to ${\bf r}_i$). Since we assume that in our setup the far-field condition is fulfilled we can limit ourselves to {\em Fraunhofer}-diffraction, i.e.,
\begin{eqnarray}\label{approx}
\hspace{-0.5cm}|{\bf r}_0-{\bf r}_i|\gg\frac{ik}{2}\,|{\boldsymbol\rho}_0|^2,
\end{eqnarray}
(though the {\em Fresnel}-integral (\ref{Ui}) can be solved numerically without this approximation). In this limit, one can carry out the integral over $\cal{A}$ in Eq.~(\ref{Ui}) to obtain the final expression for the disturbed field
\begin{eqnarray}\label{U}
\hspace{-0.5cm}\lefteqn{U({\bf r}_i,{\bf R}_j)=\frac{iA\lambda}{\pi^2}\;e^{i\frac{k}{2}\frac{2R_z^2+|{\boldsymbol\rho}_j|^2}{R_z}}e^{i\frac{k}{2}\frac{2r_z^2+|{\boldsymbol\rho}_i|^2}{r_z}}}\\
&&\cdot\frac{\sin{\left(\frac{kaR_{j_x}}{2R_z}+\frac{kar_{i_x}}{2r_z}\right)}}{R_{j_x}r_z+r_{i_x}R_z}\cdot\frac{\sin{\left(\frac{kbR_{j_y}}{2R_z}+\frac{kbr_{i_y}}{2r_z}\right)}}{R_{j_y}r_z+r_{i_y}R_z}.\nonumber
\end{eqnarray}

In Fig.~\ref{fig1}, two atoms contribute to the electric field at ${\bf r}_i$, each giving rise to a field $U({\bf r}_i,{\bf R}_j)$ of the form given in Eq.~(\ref{U}). We can thus write the total positive frequency part of the field contributing to the correlation signal at ${\bf r}_i$ as
\begin{eqnarray}\label{E}
\hspace{-0.2cm}E^{(+)}({\bf r}_i)=\frac{1}{\sqrt{2}}\,(U({\bf r}_i,{\bf R}_1)\,|g\rangle_1\langle e|+U({\bf r}_i,{\bf R}_2)\,|g\rangle_2\langle e|),
\end{eqnarray}
where the atomic operator $|g\rangle_j\langle e|$ describes the de-excitation of the $j$th atom. With $E^{(+)}({\bf r}_i)$ at hand, we can write the second order correlation function, Eq.~(\ref{G2}), in the form
\begin{eqnarray}\label{G2U}
\lefteqn{G^{(2)}({\bf r}_1,{\bf r}_2)=}&\nonumber\\
&\frac{1}{4}\left|U({\bf r}_1,{\bf R}_1)U({\bf r}_2,{\bf R}_2)+U({\bf r}_1,{\bf R}_2)U({\bf r}_2,{\bf R}_1)\right|^2.
\end{eqnarray}

In classical optics, using a coherent light source and in the limit of Fraunhofer-diffraction, we know that a rectangular aperture with opening height $a$ and width $b$ gives rise to the following classical intensity diffraction pattern at point ${\bf r}$ in the far field (see~\cite{Born:1999:a})
\begin{eqnarray}\label{I}
I({\bf r})=\left(\frac{8Ar_z}{\pi k r_x r_y R_z}\right)^2\cdot\;\sin^2{\frac{k a r_x}{2 r_z}}\cdot\;\sin^2{\frac{k b r_y}{2 r_z}},
\end{eqnarray}
where $R_z$ is the distance between the source and the aperture. From this expression we can easily recover Abbe's criterion for resolving the aperture from the intensity distribution $I({\bf r})$ in the far-field, i.e. in the Fourier plane: this is the case only if it is possible to measure the intensity diffraction pattern in the range $-2\pi<\frac{k a r_x}{2 r_z}<2\pi$ ($a\leftrightarrow b$), i.e., if we obtain the first diffraction order. In the words of Abbe: {\em we can reconstruct an image from an object if and only if the first diffraction order in the Fourier plane is at least visible}~\cite{Abbe:1873:a}. Otherwise, if we try to image an object of smaller sizes so that the diffraction pattern in the Fourier plane increases beyond that range, the image will start to blur. Note that Abbe's criterion can be used to define the resolution limit in any imaging technique (classical or non-classical). The criterion enables us in the following to compare the classical limit of $I({\bf r})$ with the two-photon correlation signal of our proposed imaging scheme.

Using Eqs.~(\ref{U}) and (\ref{G2U}), it is possible to explicitly calculate the second order correlation function $G^{(2)}({\bf r}_1,{\bf r}_2)$. For the sake of simplicity, we will fix the detectors and the source in the $x$-$z$-plane, i.e., $r_{i_y}=R_{j_y}=0$. In difference to the classical intensity pattern, we have for $G^{(2)}({\bf r}_1,{\bf r}_2)$ to determine two parameters when performing a measurement, namely ${\bf r}_1$ and ${\bf r}_2$. In addition, we have to fix the position of the two single photon emitters with respect to the aperture. The latter determines the phase shift between the different photon paths leading from either of the two emitters to the object. For example, if we choose $R_{2_x}=R_{1_x}+\pi\frac{R_z}{ka}$ and $|r_{2_x}|=r_{1_x}:=r$, we obtain
\begin{eqnarray}\label{G2final}
\hspace{-1cm}G^{(2)}({\bf r}_1,{\bf r}_2)=\left(\frac{8A^2r_z^2}{\pi^2k^2R_z^2B_\pm(r)}\right)^2\cdot\sin^2{2\frac{k a r}{2 r_z}},
\end{eqnarray}
where $B_+(r)=r^2+\frac{\pi r_z r}{ka}$ [$B_-(r)=\frac{4}{\pi}(r^2-\frac{\pi^2 r_z^2}{k^2a^2})$] holds for $r_{2_x}=+r_{1_x}$ [$r_{2_x}=-r_{1_x}$]. Comparing the modulation of the classical intensity pattern $I({\bf r})$ from Eq.~(\ref{I}) with the one obtained for the $G^{(2)}$-function in Eq.~(\ref{G2final}), one can see that the latter oscillates twice as fast as in the classical case by keeping a contrast of 100\%. Here, the additional degrees of freedom present in the $G^{(2)}$-measurement enable to select only those two-photon amplitudes which accumulate the phase $\frac{k a r}{2 r_z}$ twice as fast as in the classical case. The increase of the modulation frequency by a factor of two implies, according to the Abbe criterion, that sufficient information is available in the Fourier plane to reconstruct the aperture even if measuring only half of the range needed for the classical imaging technique.

The results found in the limit of Fresnel and Fraunhofer approximations for the case of $N=2$ emitters using $N=2$ detectors can be extended for the case of $N>2$. For example, if we consider $N=4$ single photon emitters located at positions $R_{1_x}=-\pi\frac{R_z}{ka}$, $R_{2_x}=0$, $R_{3_x}=\frac{\pi}{2}\frac{R_z}{ka}$, $R_{4_x}=\pi\frac{R_z}{ka}$ and choose for the four detectors positions $|r_{2_x}|=r$, $r_{3_x}=-r+\pi\frac{r_z}{ka}$, $r_{4_x}=r+\frac{\pi}{2}\frac{r_z}{ka}$, where again $r_{1_x}:=r$, we obtain for the fourth order correlation signal~\cite{Glauber:1963:a} the following expression
\begin{eqnarray}\label{G4}
G^{(4)}({\bf r}_1,\!{\bf r}_2,\!{\bf r}_3,\!{\bf r}_4)\propto\sin^2{4\frac{k a r}{2 r_z}}.
\end{eqnarray}
As before, we find that it is possible to image the object with sub-Rayleigh resolution. However, using four emitters and four detectors, the resolution is now enhanced by a factor of four with respect to the classical case.

\begin{figure}[t!]
\centering
\includegraphics[width=0.49\textwidth, bb=60 350 750 815, clip=true]{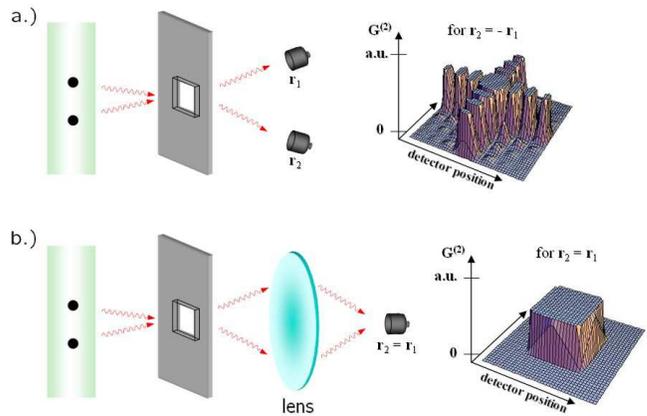}
\caption{\label{fig4} (color online) $a.$ Enhanced two-photon imaging setup employing different detector positions. $b.$ By using a lens in the Fourier plane of the object and a two-photon absorbing medium our scheme is able to reconstruct the object with sub-Rayleigh resolution without relying on a post selection mechanism.}
\end{figure}

Note that our scheme also allows to reproduce the object in the image plane of a lens placed in the Fourier plane of the object. Fig.~\ref{fig4}a recalls the initial setup under investigation in case of $N=2$. We found in Eq.~(\ref{G2final}) that an aperture with opening height $a$ generates a diffraction pattern in the Fourier plane which oscillates twice as fast when measuring $G^{(2)}({\bf r}_1,{\bf r}_2)$ than when recording $I(\bf r)$. As shown above, hereby the two detectors scan the range in the far-field at {\em different} positions, $r_{2_x}=-r_{1_x}$, so that the joint detection measurements can be performed using ordinary photon detectors (see Fig.~\ref{fig4}a). However, in order to create an image of the object in the image plane of the lens, we could also relocate the two detectors in the image plane as shown in Fig.~\ref{fig4}b. Hereby, the condition $r_{2_x}=r_{1_x}$, also compatible with $|r_{2_x}|=r_{1_x}$, i.e.~using a two-photon absorbing medium, allows us to reconstruct the object in the image plane with sub-Rayleigh resolution. Note that by using a two-photon absorbing medium the scheme does not rely on a post selection mechanism; in this way the scheme becomes relevant for lithographic applications.

Finally, we briefly outline that our method can be extended to different objects, e.g., to the case of an arbitrary grating. Therefor, let us reconsider the expression of the disturbed field $U({\bf r}_i,{\bf R}_{j_z})$ of a single rectangular aperture derived in Eq.~(\ref{U}). In the case of a grating (with $M$ slits of opening height $a$, width $b$ and slit separation $d$), each photon may pass through either of the $M$ slits before being recorded by one of the two detectors at ${\bf r}_1$ or ${\bf r}_2$. For simplicity, we restrict our calculations again to the $x$-$z$-plane. In this case, the general expression of the electric field being diffracted at a grating with $M$ slits is given by
\begin{eqnarray}\label{UN}
&&\hspace{-0.5cm}U({\bf r}_i,{\bf R}_j,M)=U({\bf r}_i,{\bf R}_j)\cdot\!\sum\limits_{n=0}^{M-1}e^{iknd\frac{R_{j_x}}{R_{j_z}}}\cdot\!\sum\limits_{n=0}^{M-1}e^{-iknd\frac{r_{i_x}}{r_z}}\nonumber\\
&&\hspace{0.8cm}=U({\bf r}_i,{\bf R}_j)\cdot\frac{1-e^{ikMd\frac{R_{j_x}}{R_{j_z}}}}{1-e^{ikd\frac{R_{j_x}}{R_{j_z}}}}\cdot\frac{1-e^{-ikMd\frac{r_{i_x}}{r_z}}}{1-e^{-ikd\frac{r_{i_x}}{r_z}}},
\end{eqnarray}
where, again, we made use of Fresnel and Fraunhofer approximations. Using this expression and by choosing $|r_{2_x}|=r_{1_x}\pm\frac{\pi r_z}{kd}$, we obtain for the second order correlation function $G^{(2)}({\bf r}_1,{\bf r}_2,M)$ for a grating with an odd number $M$ of slits
\begin{eqnarray}\label{G2N}
G^{(2)}({\bf r}_1,{\bf r}_2,M)=G^{(2)}_0({\bf r}_1,{\bf r}_2)\cdot\frac{1-\cos[k2Md\frac{r_{1_x}}{r_z}]}{1-\cos[k2d\frac{r_{1_x}}{r_z}]},
\end{eqnarray}
where $G^{(2)}_0({\bf r}_1,{\bf r}_2)$ is the second order correlation signal for a single slit aperture.

The classical expression for the intensity diffraction pattern of a grating in case of a coherent source is well known~\cite{Born:1999:a}: it consists of a sinusoidal distribution caused by the grating and an envelope function $I_0({\bf r})$ due to the diffraction by a single slit
\begin{eqnarray}\label{IN}
I({\bf r})=I_0({\bf r})\cdot\frac{1-\cos[kMd\frac{r_x}{r_z}]}{1-\cos[kd\frac{r_x}{r_z}]}.
\end{eqnarray}
Comparing Eq.~(\ref{G2N}) with (\ref{IN}), we see that both expressions can be written as a product of an envelope function due to the diffraction by a single slit ($G^{(2)}_0({\bf r}_1,{\bf r}_2)$ and $I_0({\bf r})$, respectively) and a sinusoidally oscillating function. However, in case of the $G^{(2)}$-function the modulation frequency is twice as high as in the classical case corresponding again to sub-Rayleigh resolution.

Summarizing, the proposed setup can be used to image a physical object with sub-Rayleigh resolution using uncorrelated single photon sources and linear optical tools only. In earlier articles~\cite{Agarwal:2004:a,Thiel:2007:a}, it was demonstrated how a {\em source} of single-photon emitters with a separation $d<\lambda$ can be imaged and resolved using ordinary photon detectors and joint detection techniques. In contrast, the scheme developed in this paper enables to resolve details of a {\em distinct physical object}, e.g.~an rectangular aperture or a periodical structure, impossible to resolve by classical far-field imaging techniques. We remark that the method can be implemented in various physical systems, e.g., with current ion trap technology~\cite{Maunz:2007:a,Dubin:2007:a,Dubin:2007:b}.

The authors like to thank E. Solano for helpful discussions. C.T.~and J.v.Z.~gratefully acknowledge financial support by the Staedtler foundation. G.S.A.~thanks Humboldt foundation for the continuation of the Humboldt Research Award which made this collaboration possible.


\begin{thebibliography}{50}
\bibitem{note1} We emphasize that coincident detection is not a prerequisite for our scheme. The requirement is that the two photons emitted by the two atoms are recorded by the two detectors; the exact detection time does not influence the contrast nor the resolution of the correlation signal.
\end{thebibliography}
\end{document}